\documentclass[12pt]{article}
\usepackage{graphicx}

\newcommand\pubnumber{DPF2013-130}
\newcommand\pubdate{\today}

\def\napoli{University of Arizona\\
Tucson, Arizona, USA}

\def\Title#1{\begin{center} {\Large #1 } \end{center}}
\def\Author#1{\begin{center}{ \sc #1} \end{center}}
\def\Address#1{\begin{center}{ \it #1} \end{center}}

\newcommand\pubblock{\rightline{\begin{tabular}{l} \pubnumber\\
         \pubdate  \end{tabular}}}
\newenvironment{Abstract}{\begin{quotation}  }{\end{quotation}}
\newenvironment{Presented}{\begin{quotation} \begin{center} 
             PRESENTED AT\end{center}\bigskip 
      \begin{center}\begin{large}}{\end{large}\end{center} \end{quotation}}

\def\beq{\begin{equation}}
\def\eeq#1{\label{#1}\end{equation}}
\def\eeqn{\end{equation}}

\def\beqa{\begin{eqnarray}}
\def\eeqa#1{\label{#1}\end{eqnarray}}
\def\eeqan{\end{eqnarray}}

\let\bar=\overbar

\def\Dslash{\not{\hbox{\kern-4pt $D$}}}
\def\dslash{\not{\hbox{\kern-2pt $\del$}}}

\def\msb{{\bar{\ssstyle M \kern -1pt S}}}

\begin{document}
\begin{titlepage}
\pubblock

\vfill
\Title{A Search for $t\bar{t}$ Resonances in Lepton Plus Jets Events \\
       with ATLAS using 14 fb$^{-1}$ of Proton-Proton Collisions\\ at $\sqrt{s}$ = 8 TeV}
\vfill
\Author{ Ruchika Nayyar}
\Address{\napoli}
\vfill
\begin{Abstract}
Some Beyond the Standard Model (SM) theories predict new particles that decay 
predominantly into top-antitop quark pairs. A search for top-antitop quark resonances that decay into the lepton plus jets final state is carried out with the ATLAS experiment at the LHC using 14 fb$^{-1}$ of $\sqrt{s}$ = 8 TeV proton-proton collisions. The search uses events in which the jets from hadronically decaying top quarks are resolved as well as events in which some or all of the jets from hadronically decaying top quarks are merged.
Mass exclusion limits at a 95\% CL are set for two benchmark models, one predicting leptophobic topcolor Z' 
bosons and the other predicting Randall-Sundrum Kaluza-Klein gluons.
\end{Abstract}
\vfill
\begin{Presented}
DPF 2013\\
The Meeting of the American Physical Society\\
Division of Particles and Fields\\
Santa Cruz, California, August 13-17, 2013\\
\end{Presented}
\vfill
\end{titlepage}
\def\thefootnote{\fnsymbol{footnote}}
\setcounter{footnote}{0}

\section{Introduction}

Some theories of beyond the SM physics predict the existence of new
heavy bosons that decay primarily into $t\bar{t}$ pairs.  Examples of
these theories include topcolor models~\cite{zprime_model}, chiral
color models~\cite{chiral_model} and Randall-Sundrum models with
warped extra dimensions~\cite{rs_model}. The ATLAS experiment has
searched for the production of top quark pair resonances produced in
14.3 fb$^{-1}$ of proton-proton (pp) collisions at a center-of-mass
energy of 8 TeV~\cite{conf_note}. The search is carried out in the lepton plus jets
decay channel where one $W$ boson from a top quark decays leptonically
and the other decays hadronically. The $t\bar{t}$ invariant mass
spectrum is then tested for any local excess of events. The data is
tested against two particular theoretical models that differ in the
width of the $t\bar{t}$ resonance with respect to detector mass
resolution.  A benchmark model that produces narrow width $t\bar{t}$
resonance is that of the topcolor, leptophobic $Z'$. A benchmark model
that produces wide width $t\bar{t}$ resonance is the Kaluza Klein
gluon that arise in Randall$-$Sundrum models with an extra dimension
with a warped geometry and where the entire Standard Model fields and
matter can propagate in all five dimensions (bulk).


\section{Data and Monte Carlo Samples}

The data sample used in this search was collected with single lepton triggers with exactly one lepton in the final state.
The integrated luminosities for the electron and muon data sets collected during 2012 running are $14.3\pm 0.5$ fb$^{-1}$ and $14.2 \pm 0.5 fb^{-1}$. 


SM $t\bar{t}$ production is the primary irreducible background which is modeled using the MC@NLO~\cite{mcatnlo} generator, 
Herwig~\cite{herwig} for parton showering and hadronization and Jimmy~\cite{jimmy} for modeling the multiple parton scattering. 
The CT10~\cite{ct10} parton distribution functions (PDFs) are used and the top quark mass is set to 172.5 GeV. Single top quark production 
in the s-channel and production with an associated W (Wt) are modeled via 
MC@NLO/Herwig/Jimmy~\cite{single_top} as above. Production in the t-channel is modeled using the AcerMC~\cite{acermc} generator and 
Pythia~\cite{pythia} for parton showering and hadronization. Samples for leptonic decays of W and Z bosons, including those to $\tau$, 
accompanied by jets are generated with Alpgen~\cite{alpgen} with up to five extra 
final state partons at leading order without virtual corrections. Specific W boson plus heavy flavor processes 
are generated separately with Alpgen and double counting of the heavy flavor contributions is removed from the W plus light quark jets samples. 
The Z+jets production is modeled using Alpgen and includes contributions from the interference between photon and Z boson exchanges and events 
are required to have a dilepton invariant mass $40 < m_{ll} <2000~GeV$. The overlap removal for heavy flavors is done as in the case of W+jets samples.
The multijet background is estimated using data. 

Signal samples of topcolor $Z'$ are modeled using the SSM $Z' \to t\bar{t}$ process as implemented in Pythia 
with MSTW2008LO PDFs~\cite{pdf}. A K-factor of 1.3~\cite{kfactor} is applied to account for NLO effects. Signal samples of Randall-Sundrum KK gluons 
were generated via Madgraph~\cite{madgraph} and then hadronized using Pythia. 



\section{Event Selection}

Events are required to pass a high-$p_T$ single electron or muon lepton trigger with exactly one lepton in the final state. The events must have a reconstructed primary vertex with at least five tracks with $p_T > 0.4~GeV$. In the $e+ jets$ channel, $E_{T}^{miss}$ must be larger than 30 GeV and the transverse mass $(m_T)$ larger than 30 GeV.
Whereas In the $\mu+jets$ channel, $E_{T}^{miss}$ must be larger than 20 GeV and the $E_{T}^{miss} + m_T$ larger than 60 GeV. 

The selected events are then classified as either originating from boosted or resolved hadronic top quark decays. The boosted sample consists of events with at least one 
small-radius $(\Delta R=0.4)$ jet and at least one large-radius $(\Delta R=1.0)$ jet. The small-radius jet is the highest-$p_T$ jet satisfying $\Delta R(l,j) <1.5$ where $l$ is selected lepton. The large-radius jet must satisfy $m_{jet}>100~GeV$, first $k_T$ splitting scale $d_{12} > 40~GeV$, $\Delta R(jet, j_{sel}) > 1.5$ and $\Delta\phi(jet,l) >2.3$. Here $l$ is the selected lepton and $j_{sel}$ is the small-radius selected jet. 

For the resolved selection, the event is required to have at least  three small-radius jets. Furthermore it is required that one of those jets has mass $> 60~GeV$ or there is another, fourth small-radius jet satisfying $p_T > 25~GeV$, $|\eta| < 2.5$ and $|JVF| > 0.5$. Finally it is required that at least one of the selected small-radius jet should be b-tagged. 

\section{Event Reconstruction}
The invariant mass of the top quark pairs is reconstructed using the four-momenta of the physics objects in the 
event. For the semi-leptonically decaying top quark, in both the resolved and the boosted selections, the $p_z$ is computed by 
imposing an on-shell $W$ boson mass constraint on the lepton plus $E_{T}^{miss}$ system. If only one real solution to $p_z$ exists, this is 
used. If two real solutions exist, the solution with the smallest $|p_z|$ is chosen. If no real solution is found, the $E_{T}^{miss}$ is rescaled 
and rotated by applying a minimum variation to find a real solution. 

\begin{figure}[htb]
\centering
\includegraphics[height=1.5in]{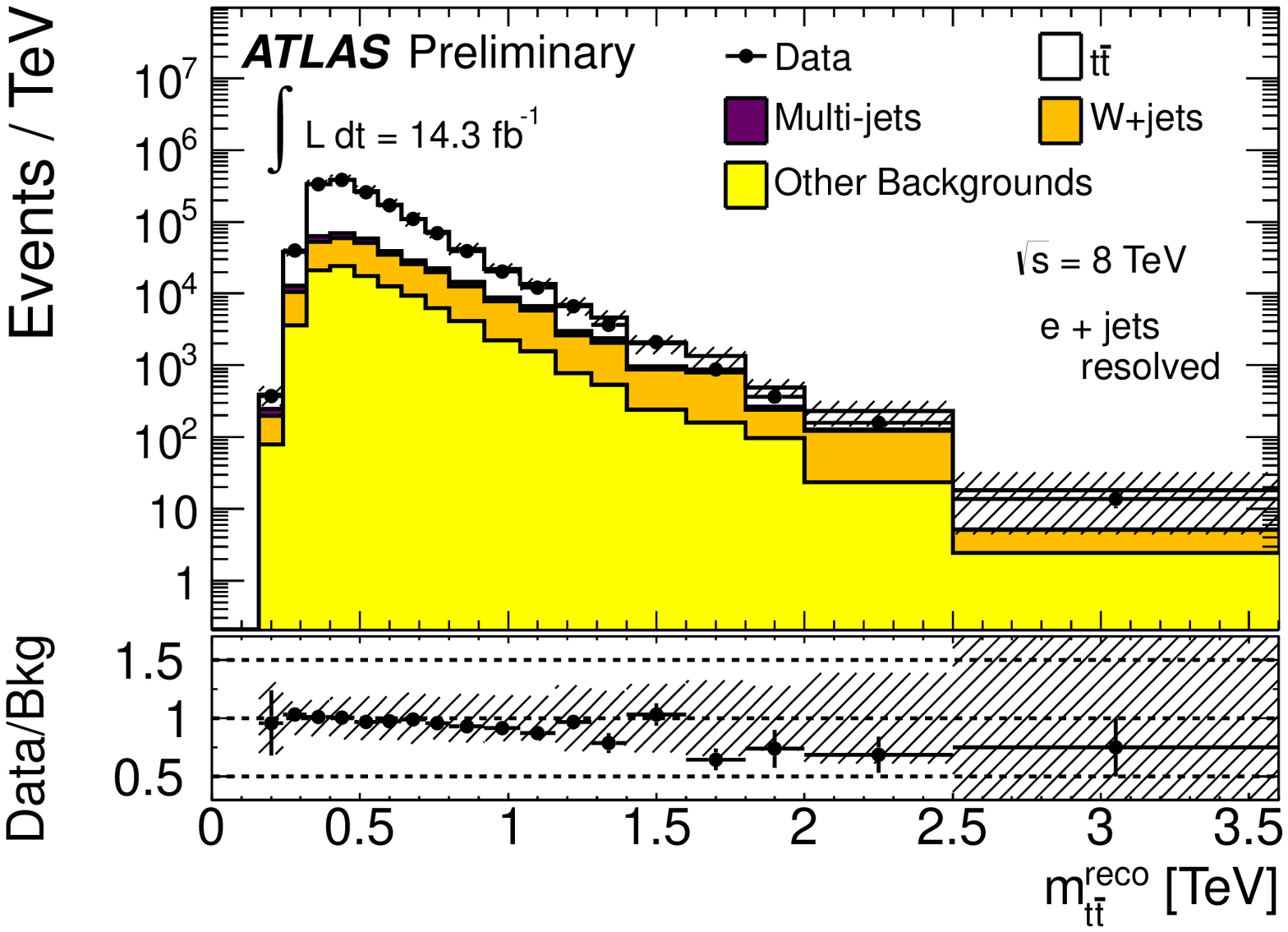}
\includegraphics[height=1.5in]{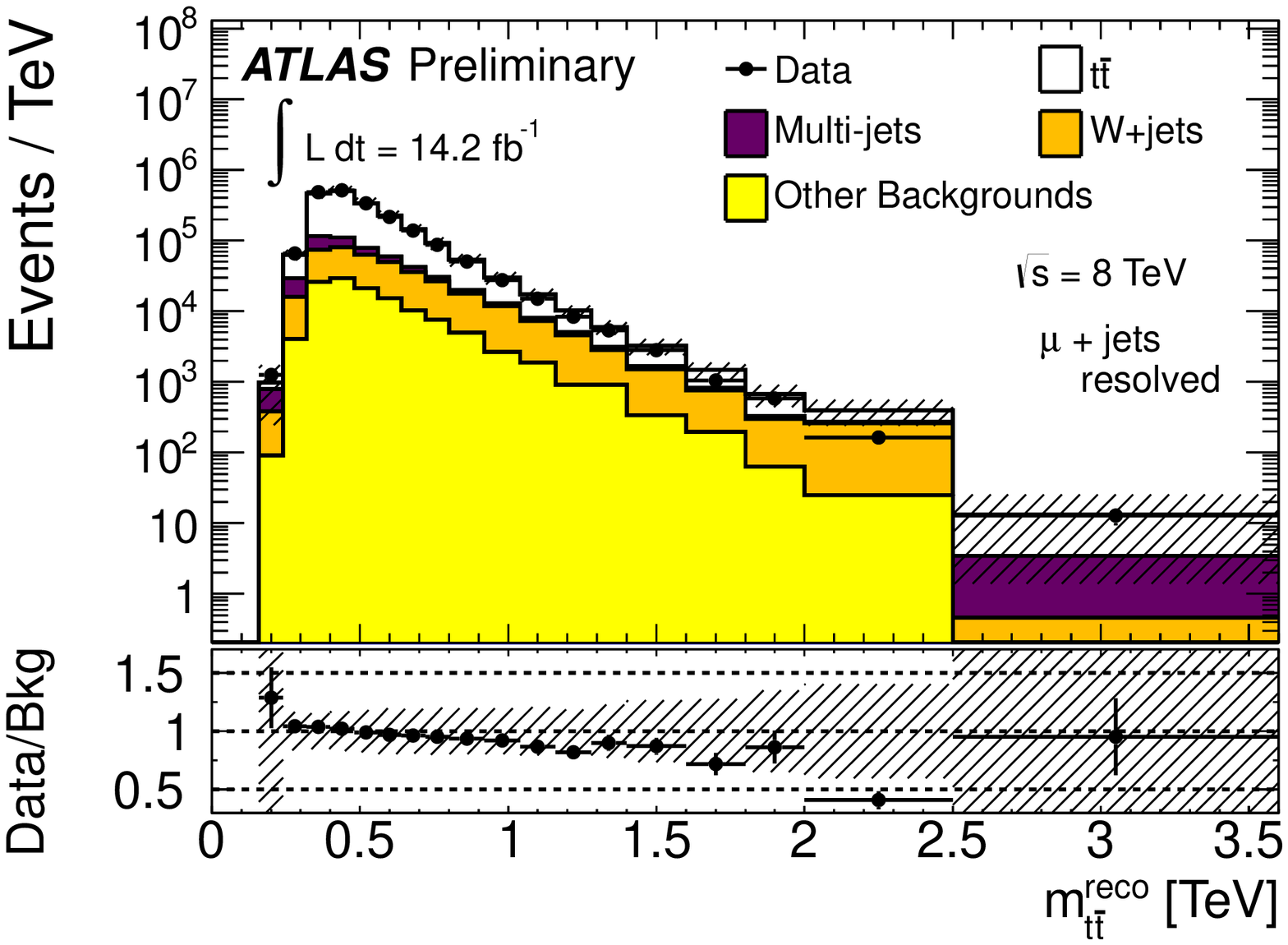}\\
\includegraphics[height=1.5in]{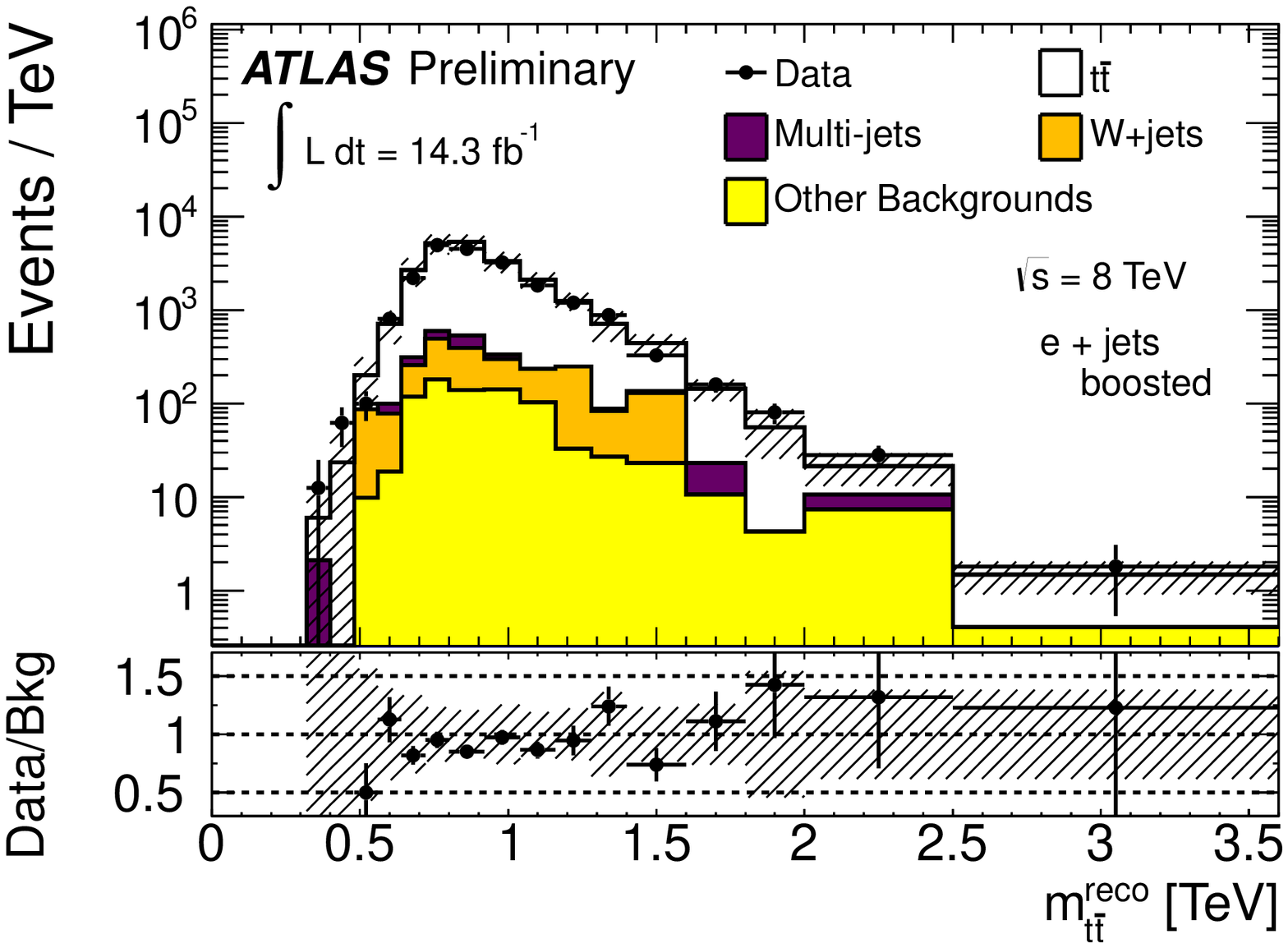}
\includegraphics[height=1.5in]{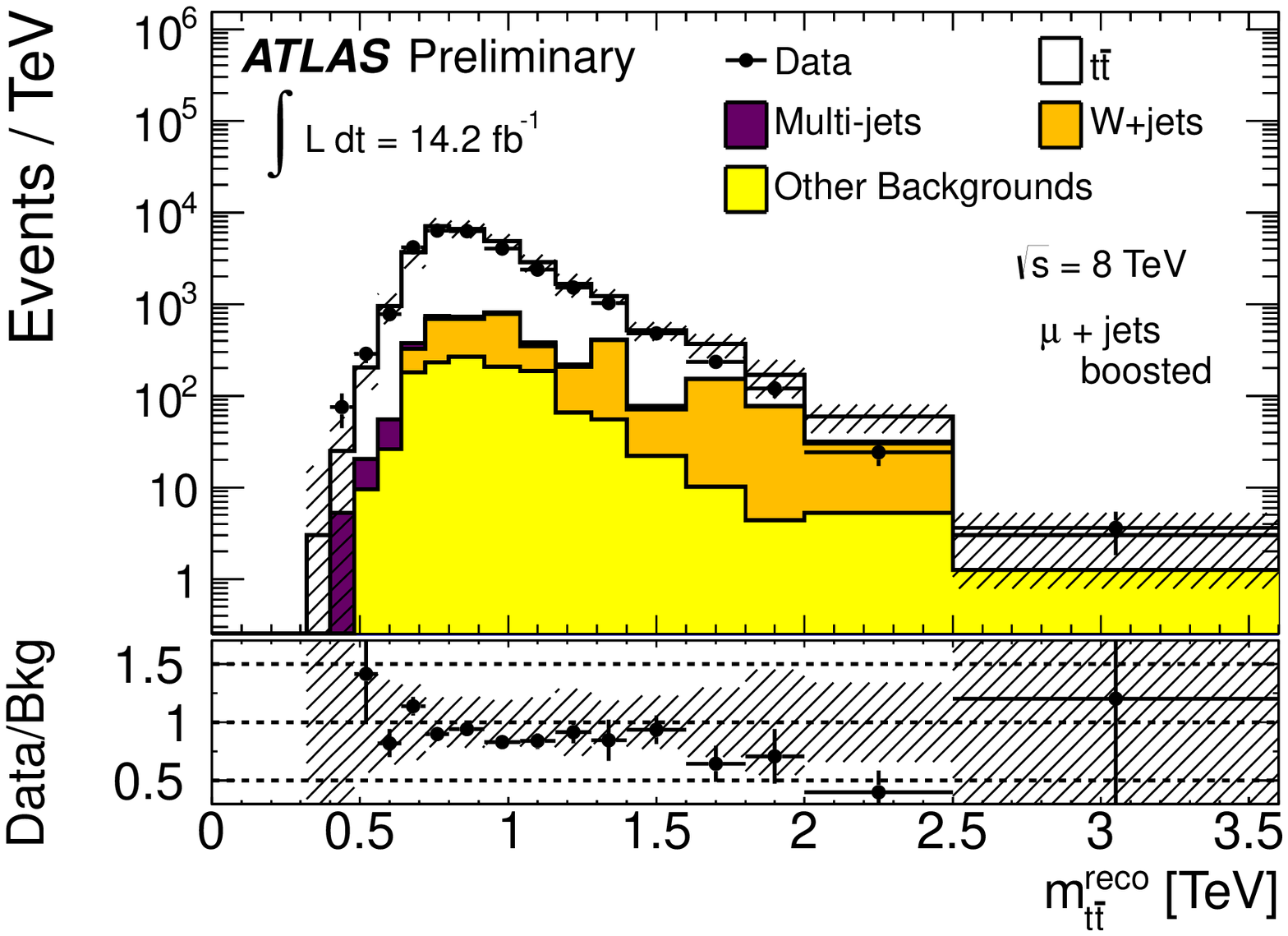}\\
\caption{The $m_{t\bar{t}}^{reco}$ distribution for resolved selection (top row) and 
boosted selection (bottom row).}
\label{mass}
\end{figure}

\begin{itemize}
\item For resolved reconstruction regime, a $\chi^2$ algorithm is used to select the best assignment of jets to the
hadronically and semi-leptonically decaying top quarks. The algorithm uses the reconstructed top 
quark and $W$ boson masses as constraints. All possible permutations for four or more jets are tried and
the permutation with the lowest $\chi^2$ is used to compute the $m_{tt}^{reco}$. 
\item For the boosted reconstruction, there is no ambiguity in the assignment of jets. The hadronically
decaying top quark four-momentum is taken to be that of the large-radius jet, while the semi-leptonically
decaying top quark four-momentum is formed from the neutrino solution from the $W$ boson mass constraint, 
the high $p_T$ lepton and the selected small-radius jet.
\end{itemize}
The reconstructed spectra for $t\bar{t}$ for each selection and channel is shown in Fig.~\ref{mass}. A good agreement is observed 
between data and the total background expectation. 
\section{Limits}
\begin{figure}[htb]
\centering
\includegraphics[height=2.0in]{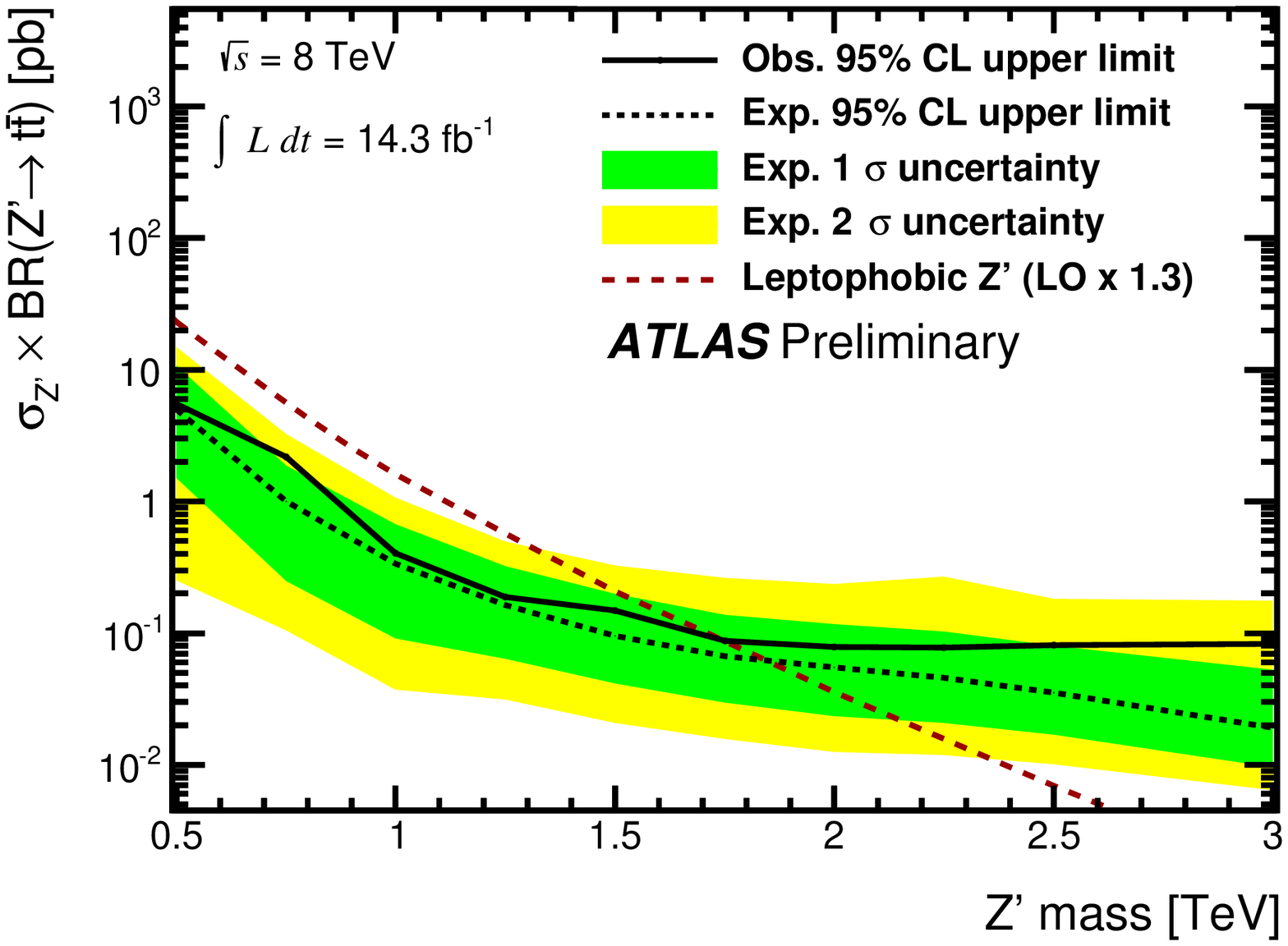}
\includegraphics[height=2.0in]{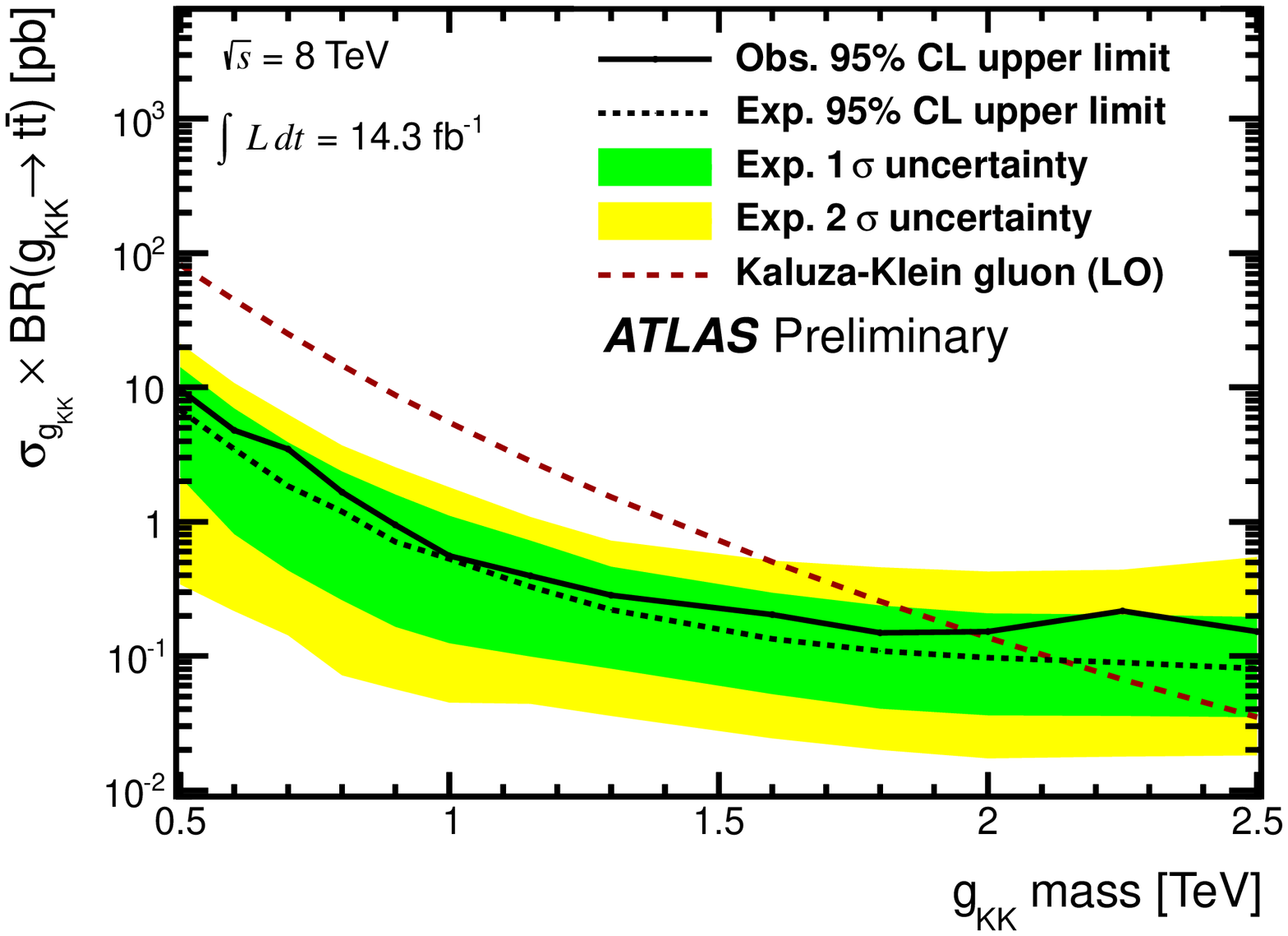}
\caption{Expected and observed upper cross section limits times the branching ratio on $Z'$ (left)
and KKg(right).}
\label{limits}
\end{figure}

After the reconstruction of the $t\bar{t}$ spectra, the data and expected background distributions are
compared to search for hints of new physics using BumpHunter~\cite{bumphunter}, which is a hypothesis testing tool
that searches for local excesses or deficits in the data compared to the expected background, taking the
look-elsewhere effect into account over the full mass spectrum. The analysis considers various sources 
of systematics that affects the shape and normalization of the spectra.
The dominant source of systematics arise from $t\bar{t}$
normalization, jet energy scale for small-radius and large-radius
jets, the b-tagging efficiency and PDF. After accounting for the
systematic uncertainties, no significant deviation from the expected
background is found.  Hence upper limits are set on the cross section
times branching ratio of the $Z'$ and KKg benchmark models using a
Bayesian technique.  In the combination, the four statistically
uncorrelated spectra are used, corresponding to boosted and resolved
selections as well as $e+jets$ and $\mu+jets$ channel.  Using the
combined upper cross section limits, a leptophobic topcolor $Z'$ boson
(KK gluon) with mass between 0.5 and 1.8 TeV (0.7 and 2.0 TeV) is
excluded at 95\% CL.  The upper cross section limits with systematic
and statistical uncertainties are given for the two benchmark models
for the combination of the two selections (see Fig.~\ref{limits}).

\end{document}